# NourID+:
# A Digital Energy Identity Framework for Efficient Subsidy Allocation in Morocco

Fatima Zahra Iguenfer, Younes Lamhamedi Cherradi, Nada Belkhayat, and Hiba Jebbar
*School of Sciences and Engineering*
*Al Akhawayn University*
Ifrane 5300, Morocco
{f.iguenfer, y.lamhamedicherradi, n.belkhayat, h.jebbar}@aui.ma

*Abstract*—We introduce NourID+, a digital energy identity framework that addresses Morocco's need for trusted energy subsidy allocation through authenticated digital identity integration. NourID+ creates a strong foundation for future subsidy programs by unifying three government-issued and digitalized credentials: Moroccan national identity cards (CIN), cadastral plans, and property ownership certificates are transformed into unique digital energy IDs (DE-IDs) that map authenticated identities with specific properties and their energy consumption patterns. The system supports three property ownership profiles: farmers (landowners), entrepreneurs (factory or company owners), and households (house owners), as energy consumption is directly related to land ownership. NourID+ provides dual access through a government portal allowing officials to process DE-ID generation requests, as well as a citizen portal for DE-ID usage and energy monitoring. Our framework supports CIN upload with facial biometric matching, automated property retrieval through government APIs, and government officer approval workflow for DE-ID generation. After evaluation of the system, we demonstrate a reduction in verification time from weeks to minutes, with 98% accuracy of document validation. The proposed solution allows for targeted subsidy allocation of electricity based on actual consumption needs rather than estimations, potentially improving the efficiency of Morocco's significant energy subsidy expenditure.

*Index Terms*—Digital Identification, Digital Governance, AI-Driven Energy Monitoring, Sustainable Development, Subsidies Allocation.

## I. INTRODUCTION

Since the 1930s, Morocco has invested in energy and other subsidies for the purpose of protecting vulnerable population groups and to promote domestic industries. However, and starting 2007, the country faced a rising fiscal deficit that was beyond the government's control. Food and energy subsidies reached 6.5% of GDP in 2012, with 70% going to energy products [1]. In 2022, the budgeted subsidy bill was approximately 42 billion MAD (about 4 billion USD), with the government aiming to further cut the fiscal deficit in the next few years [2]. Furthermore, in the same year, Morocco's energy subsidies amounted to 5.8% of its country's gross domestic product, spending $7.6 billion on energy subsidies primarily based on fossil fuels such as gasoline, gas and butane [3] while the household electricity tariffs have remained broadly cost reflective.

Despite Morocco's commendable efforts to achieve near-universal electricity access, rising from 18% in 1995 to 99.5% in 2017 through the Program d'É´lectrification Rurale Ge´ne´ralise´e (PERG) [4], a significant portion of the population continues to struggle with energy poverty. Recent studies indicate that approximately 14% of Moroccan households are energy poor, predominantly large families living in rural areas with limited income and education levels [5]. Although electricity tariffs are generally cost-reflective, the Moroccan government continues to allocate substantial funds to stabilize prices, especially in response to increasing global energy costs. In 2024, for example, 4 billion MAD were allocated to the National Office of Electricity and Drinking Water (ONEE) to prevent potential price increases for consumers [6].

However, stabilizing electricity prices across the board is fiscally demanding, and without a mechanism to identify and prioritize the most vulnerable households, such spending lacks efficiency. Redirecting this public expenditure toward those most in need requires a system capable of accurate, real-time, property-linked energy tracking, and targeted subsidy allocation. Yet, effective allocation and monitoring of subsidies is still hindered by fragmented records and the absence of digitalized energy tracking platforms. This underscores the urgent need for digital transformation. In line with this vision, the Moroccan government has launched the ambitious "Digital 2030" strategy, which aims to position the country as a leader in digital innovation on the African continent, driving economic growth, improving public services and improving the quality of life of citizens [7].

Although various studies have explored energy monitoring and optimization approaches using IoT technologies [8], machine learning [9], and web-based systems [10], a critical gap persists in the integration of authenticated digital identities with energy consumption tracking into a unified verifiable system. Most existing solutions focus on monitoring usage or optimizing performance, but fail to tie energy consumption to legally recognized property and identity data.

To bridge this gap, we propose **NourID+**, a digital energy identity framework that consolidates three already digitalized government issued credentials: the CIN, cadastral plans and property ownership certificates. This integration allows for the generation of a unique DE-ID that links specific land parcels, not just individuals, to verified energy usage. In doing so, NourID+ allows for more accurate, transparent, and equitable allocation of electricity subsidies, fully aligned with the objectives of the Moroccan strategy "Digital 2030".

Moreover, NourID+ aligns with Morocco's recent structural reforms in public utility governance. The promulgation of Law n° 83-21 on 12 July 2023 established the legal framework for Sociétés Régionales Multiservices (SRM), regionally operated entities tasked with managing water, electricity, sanitation and public lighting at the local level [11], [12]. The first such entity, the Société Régionale Multiservices Casablanca-Settat (SRM-CS), officially began operations on October 1, 2024 [13]. SRM-CS replaces former operators including LYDEC, RADEEJ, RADEEC, and the regional distribution units of ONEE, centralizing utility governance for the entire Casablanca-Settat region [14]. In future work, our goal is to collaborate with SRM to integrate NourID+ with real-time consumption data through government APIs, allowing full deployment of our digital energy identity framework and contributing to Morocco's broader effort to modernize public utility management.

The main contributions of this paper are summarized as follows:
1) We introduce a framework that merges three government-issued digital credentials, CIN, cadastral plans, and property ownership certificates, into a unique DE-ID that identifies land parcels, not just individuals.
2) We design a verifiable digital identity system to help the Moroccan government allocate electricity subsidies accurately based on authenticated property-level consumption data.
3) We propose a future collaboration with the SRM to integrate NourID+ with real consumption data and support the deprivatization of Morocco's energy distribution sector.
4) We provide a replicable system for digital energy governance that can be adopted by other developing regions.

The remainder of this paper is organized as follows. Section II reviews related work. Section III presents the NourID+ system. Section IV discusses our implementation results, including system testing results and energy consumption monitoring capabilities. Finally, Section V concludes the paper with a summary of our contributions and directions for future research.

## II. RELATED WORK

Recent advances in digital energy monitoring systems have significantly transformed the way we track, analyze, and optimize energy consumption in various sectors. The work [15] introduces a digital energy monitoring system that reduces human error by accurately measuring electricity from portable sources, offering a user-friendly dashboard accessible through individual user accounts for real-time monitoring and consumption analysis. Similarly, [16] presents a web-based real-time monitoring system for electrical energy consumption in residences powered by micro-CPV systems, providing high accuracy with an error rate of 0.6% in tracking electrical power consumption. The paper [17] advances this field by presenting a GSM-based digital energy monitoring device incorporating fuzzy-based models to simulate the characteristics of current and voltage sensors, while [18] builds smart energy meters that monitor household consumption while integrating overload protection and power theft control through microcontroller-based systems.

Beyond basic monitoring, advanced energy management approaches have emerged. The study [19] introduces an Intrusive Energy Management (IEM) approach within the multilayer IoT architecture designed to identify user activities through sensor data. These Smart Energy Management Systems (SEMS) promote energy efficiency and sustainability through advanced control techniques, enabling users to monitor energy usage, improve voltage stability, and support grid operators through load scheduling and demand-side management strategies.

However, most existing systems focus on real-time monitoring, appliance-level optimization, or isolated subsidy mechanisms, with few exploring the use of these technologies to generate unique identifiers linked to specific properties for systematic energy tracking. This reveals a clear gap in comprehensive frameworks that address both the technical and socioeconomic dimensions of energy distribution. Without verifiable digital identity tied to property-level consumption, implementing data-driven policies for subsidy targeting and infrastructure planning becomes difficult, particularly in regions where accurate tracking is critical for sustainability.

To address these limitations, several African countries are pioneering the connection of digital ID systems to public utilities. Ghana has started to link smart electricity meters to national biometric IDs (GhanaCard) to ensure accountability in household energy consumption, each new meter being associated with verified identification and address to prevent fraud and allow for targeted energy policy [20]. Ethiopia's "Fayda" ID program serves as the foundational ID for all government services, including land registration and social assistance, creating a nationwide link between identity and service delivery [21]. On the other hand, Rwanda's blockchain-enabled land registry integrates digital ownership records with citizen identity through its Land Administration Information System (LAIS) and the Irembo portal, allowing secure and tamper-proof verification for service access [22]. Finally, countries such as Nigeria exemplify this trend by reformatting universal energy subsidies through targeted digital cash transfers. The Nigerian government's safety net program now biometrically verifies each beneficiary's national identification number (NIN) or bank verification number (BVN) before payment, ensuring that only eligible poor households receive relief [23].

These developments reflect a growing recognition that verified identity and ownership data must converge to provide efficient, secure, and inclusive public services. The integration of digital identity with energy systems is gaining traction globally, with blockchain-based energy identifiers developed for enterprise-level green certification in China [24]. However, such approaches remain limited in Africa and are rarely applied in residential or agricultural settings. Our work fills this gap by proposing a verifiable DE-ID that integrates national identity, authenticated property ownership, and energy usage through platforms such as the Modular Open Source Identity Platform (MOSIP) [25], building on the foundation established by these emerging African digital identity initiatives.

## III. System Architecture

### A. System Overview

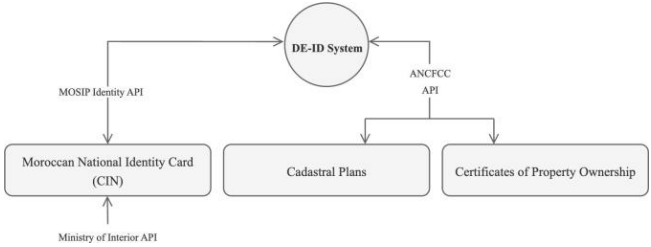

Fig. 1. Overview of the NourID+ framework showing the integration of three government credentials (CIN, cadastral plans, property certificates) into unique Digital Energy IDs for subsidy allocation

NourID+ creates a comprehensive digital energy identity framework that streamlines the entire process, from citizen registration to monitoring of energy consumption. As illustrated in Fig. 1, our system creates unique DE-IDs by systematically combining three key government issued credentials: the Moroccan CIN, cadastral plans and certificates of property ownership. However, unlike traditional systems that require manual document submission, our framework automatically retrieves and validates these credentials through secure API integrations with the Ministry of Interior and the National Agency for Land Conservation, Cadastre, and Cartography (ANCFCC).

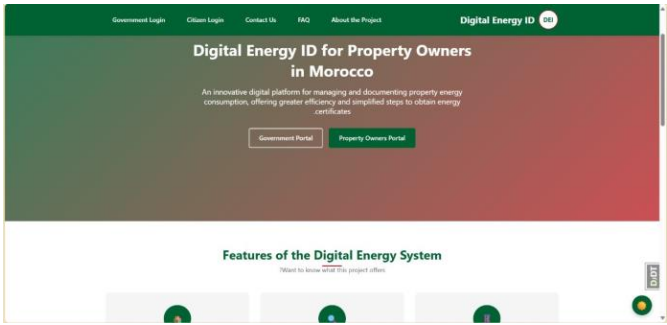

Fig. 2. Dual-portal architecture for citizen registration flow and government officer approval workflow for DE-ID generation

Fig. 2 above shows how the framework operates through two primary portals: a citizen-facing platform where users can create accounts, verify their identity, select properties, and monitor energy consumption; and a government portal where SRM officers review and approve DE-ID generation requests. This dual-portal approach ensures both citizen accessibility and government oversight while maintaining security and compliance with Morocco's digital governance standards.

The innovation of the system lies in its property-centric approach to energy identity. Rather than simply linking individuals with energy consumption, NourID+ creates verified associations between authenticated citizens and their legally owned properties, allowing for precise subsidy allocation based on actual property-level consumption data.

### B. User Registration and Identity Verification

The user journey begins when the citizen creates an account on the platform providing essential information including full name, email address, phone number, and secure password. This initial registration establishes the user's digital presence within the system while maintaining minimal data collection principles. Upon successful creation and log-in, users enter the identity verification phase, which represents the cornerstone of the NourID+ security framework. As shown in Fig. 3, the system presents an intuitive interface in which users upload their CIN and perform facial biometric scanning using the integrated biometric capture system of the platform. The facial recognition component uses an advanced computer vision model with accuracy 98% to capture high-quality biometric data.

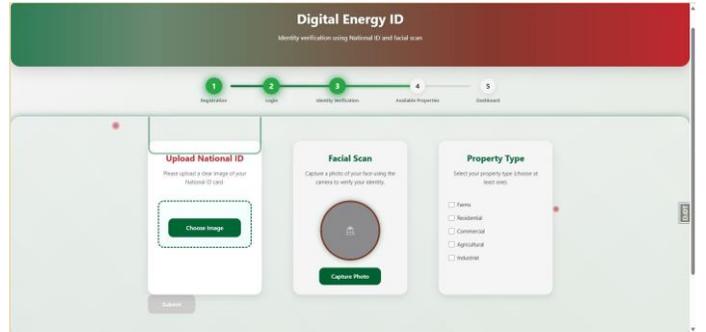

Fig. 3. User interface for CIN upload and facial biometric verification

### C. Government API Integration and Verification

Following the capture of user credentials and biometric data, NourID+ starts a secure verification process through integration with the Ministry of Interior's national identity database. The system transmits the CIN information uploaded and the facial biometric data captured to the Ministry of Interior API, which performs a comprehensive validation against official government records. This API then returns the verification status along with the confirmed identity details, which are later securely stored within the NourID+ system. After successful identity verification, the system automatically queries the Ministry of Interior's Property Registry API to retrieve a comprehensive list of all properties legally associated with the user's CIN.

### D. Property Selection and Document Retrieval

The verified property information is then displayed to the users, as illustrated in Fig. 4. A comprehensive view of all properties associated with the user's identity is displayed, including property types, locations, area, and other relevant information. Users can select one or multiple properties for energy monitoring and subsidy eligibility, providing flexibility for property owners with diverse portfolios.

For each selected property, NourID+ automatically initiates document retrieval through the ANCFCC API, fetching essential documentation, including cadastral plans and certificates of ownership. The cadastral plans provide precise spatial and geographic boundary information, while the ownership certificates establish legal property rights.

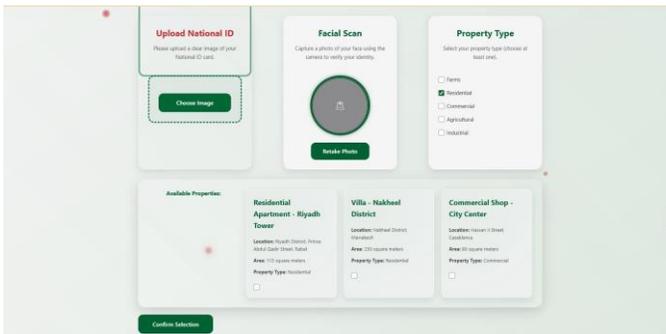

Fig. 4. Property selection interface displaying user's verified properties retrieved through government APIs

### E. Government Officer Approval Workflow

Another distinctive feature of NourID+ is its integration of government oversight through the SRM officer approval workflow. Upon completion of document validation, the system generates a formal DE-ID issuance request that is submitted to designated government officers within the SRM framework. Fig. 5 shows a list of requests as seen by a government officer. Each request includes verified user identity information, validated property documents, and a detailed summary of the DE-ID issuance requested. Officers can approve or reject requests based on policy guidelines and document validation results.

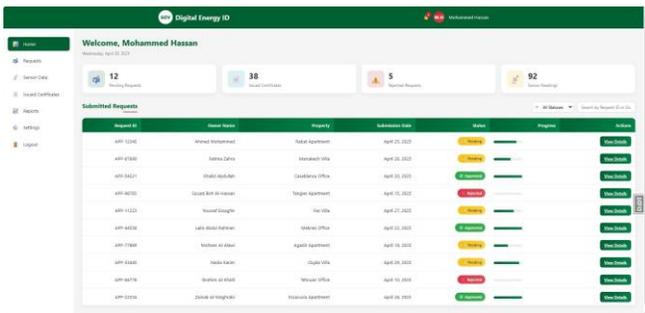

Fig. 5. SRM officer portal for reviewing and approving DE-ID requests

### F. Digital Energy ID Generation and QR Code Issuance

Upon officer approval, NourID+ generates unique DE-IDs for each approved property. The DE-ID generation process synthesizes multiple data elements including the user's verified CIN, official property identifiers from cadastral documents, and system-generated unique identifiers to create a comprehensive digital identity tied to specific properties. As illustrated in Fig. 6, each DE-ID encodes essential information while maintaining privacy and security standards. The system generates the corresponding QR codes that serve as practical access mechanisms for energy-related services. These QR codes incorporate appropriate error correction levels to ensure readability even in challenging environmental conditions commonly encountered in rural areas.

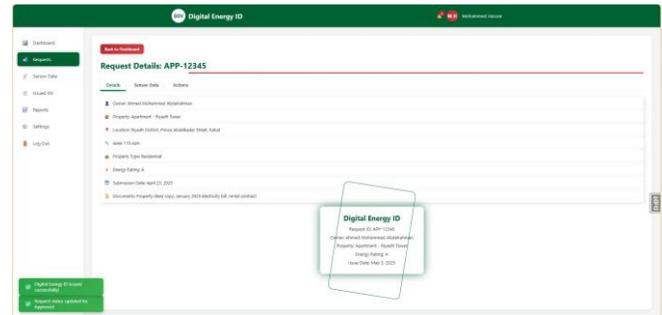

Fig. 6. Digital Energy ID generation interface creating QR codes for property-specific energy service access

### G. Energy Consumption Monitoring and Dashboard Analytics

The final component of the NourID+ framework provides a comprehensive energy consumption monitoring through integration with official energy provider databases. The system interfaces directly with the existing utility infrastructure, particularly the SRM metering databases, to access real-time electricity consumption data linked to each DE-ID.

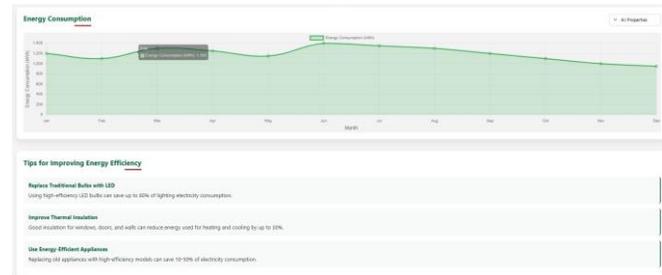

Fig. 7. Energy consumption dashboard providing multi-temporal analytics and AI-powered consumption forecasting

Users access their energy consumption information through personalized dashboards within the citizen portal, as shown in Fig. 7. These dashboards provide detailed analytics across multiple time frames, including daily, weekly, monthly, and yearly consumption patterns.

Dashboard analytics include not only trend analysis, but also forecasting capabilities thanks to a dedicated AI module. We use a time-series machine learning model based on XGBoost, which takes advantage of lag features, rolling averages, and seasonal patterns extracted from historical energy consumption data to predict future usage. This comprehensive monitoring system enables citizens and government officials to make data-driven decisions regarding energy usage and subsidy allocation, supporting Morocco's broader sustainability and efficiency objectives under the Digital 2030 strategy.

## IV. RESULTS AND DISCUSSIONS

### A. System Implementation and Testing

The NourID+ system architecture has been designed following the framework outlined in Section III. The modular components of the system have been individually prototyped and validated to demonstrate technical feasibility. For a complete end-to-end implementation, strategic partnerships with SRM as well as relevant government agencies would be necessary to access the required APIs, including the Ministry of Interior and ANCFCC. To evaluate the performance of the system, we conducted simulations with synthetic workflows that represent the entire verification process. These simulations included representative user personas from our target demographics (farmers, homeowners, and entrepreneurs). The simulation results indicated that a fully implemented system can complete the verification workflow in a few minutes instead of days or even weeks typically required by current paper-based processes.

### B. Energy Consumption Monitoring

A key innovation in the NourID+ system is that, upon successful DE-ID generation, the system creates a personalized digital space accessible through our web platform. This dashboard provides users with detailed visibility into their energy consumption patterns across multiple time frames (day, week, month, year), as shown in Fig. 7.

Although the production system is designed to collect real-time data from existing utility infrastructure through SRM APIs, our current implementation utilizes synthetic data to demonstrate functionality. These simulated datasets were generated based on typical Moroccan energy consumption patterns, accounting for seasonal variations, different types of property (residential, agricultural, commercial), and occupancy scenarios including both property owners and tenants.

### C. Comparative Context: African Digital ID and Energy Subsidy Systems

African countries are rapidly building digital ID backbones for public services, but few have explicitly tied these to energy subsidies. For example, South Africa's Smart ID (national biometric identification card) is the basis for its social grant system [26]. On the other hand, Nigeria is linking its National Identification Number (NIN) to targeted subsidies: the Agriculture Ministry now maps the NIN of each farmer (and biometrics) to land and a multi-wallet payment card for input subsidies [27]. To this end, the World Bank notes that up-to-date social registries with verified digital identities greatly improves pro-poor targeting [28], and the Nigerian case shows that biometrics can enforce that only eligible people receive aid [23].

In contrast, and to the best of our knowledge, no African system explicitly links a national digital ID to household electricity accounts in a way that enables real-time consumption monitoring, property-based verification, and automated subsidy allocation within a unified platform. In this sense, Morocco's proposed NourID+ framework is notably novel. The broader lesson from existing initiatives is that digital ID–linked subsidies can significantly improve efficiency and equity when built on comprehensive, interoperable databases, but they must also safeguard inclusion and privacy. Achieving this balance is the main objective of the NourID+ solution.

## V. CONCLUSION

This paper introduces NourID+, a digital energy identity framework that merges Morocco's national identity framework with cadastral and property data to facilitate the selective distribution of electricity subsidies. By consolidating three official credentials already digitalized by the Moroccan government, the National Identity Card, cadastral plans, and property ownership documents, into a verifiable DE-ID. Our system offers a new approach to a long-standing policy issue in Morocco: the ineffective allocation of subsidies resulting from disjointed records, inadequate digital infrastructure, and the prevalence of privatized utility providers. The dual-portal system helps government personnel authenticate DE-ID requests while allowing citizens to use personalized dashboards to visualize energy consumption at the property level. Simulated workflow evaluations suggest that the system can decrease verification time from weeks to minutes and eliminate duplication and fraud in subsidy distribution.

Morocco's recent institutional reforms through Law n° 83-21 have enhanced the framework's potential by forming Sociétés Régionales Multiservices to unify and manage utility distribution. Collaborating strategically with SRM would allow NourID+ to connect to real-time utility APIs and move from simulation to active deployment. This partnership marks an important step towards deprivatizing the energy distribution and aligning with the "Digital 2030" national initiative. We also designed NourID+ to be expandable to other regions within Morocco and adaptable to similar African areas experiencing similar issues.

## VI. ACKNOWLEDGMENTS

We acknowledge the Upanzi Network and MicroSave Consulting (MSC) for organizing the Digital ID Hackathon Africa 2025, where NourID+ received the first prize for the runners-up award. This hackathon challenged African university students to develop high-impact digital ID use cases across various sectors, with final presentations at the ID4Africa 2025 General Meeting in Addis Ababa, Ethiopia. We thank Carnegie Mellon University and the ID4Africa community for providing a platform to showcase our digital energy identity framework and for the valuable feedback that continues to shape our work toward Morocco's Digital 2030 objectives.